\documentclass[aps,prb,twocolumn,superscriptaddress,oneside,floatfix,amsmath,showpacs,amssymb]{revtex4-1}
\usepackage{graphicx}
\usepackage{dcolumn}
\usepackage{bm}
\usepackage[usenames]{color}

\begin{document}

\newcommand{\U}{URu$_2$Si$_2$}
\newcommand{\ie}{\textit{i.e.}}
\newcommand{\eg}{\textit{e.g.}}
\newcommand{\etal}{\textit{et al.}}


\title{Hybridization gap and anisotropic far-infrared optical conductivity of URu$_2$Si$_2$}


\author{J.~Levallois}
\affiliation{D\'{e}partement de Physique de la Mati\`{e}re Condens\'{e}e, Universit\'{e} de Gen\`{e}ve, CH-1211 Gen\`{e}ve 4, Switzerland}

\author{F.~L\'{e}vy-Bertrand}
\affiliation{D\'{e}partement de Physique de la Mati\`{e}re Condens\'{e}e, Universit\'{e} de Gen\`{e}ve, CH-1211 Gen\`{e}ve 4, Switzerland} \affiliation{Institut N\'{e}el, CNRS et Universit\'{e} Joseph Fourier, BP 166, F-38042 Grenoble Cedex 9, France}

\author{M.K.~Tran}
\affiliation{D\'{e}partement de Physique de la Mati\`{e}re Condens\'{e}e, Universit\'{e} de Gen\`{e}ve, CH-1211 Gen\`{e}ve 4, Switzerland}

\author{D.~Stricker}
\affiliation{D\'{e}partement de Physique de la Mati\`{e}re Condens\'{e}e, Universit\'{e} de Gen\`{e}ve, CH-1211 Gen\`{e}ve 4, Switzerland}

\author{J.A.~Mydosh}
\affiliation{Kamerlingh Onnes Laboratory, Leiden University, 2300RA Leiden, The Netherlands}

\author{Y.-K.~Huang}
\affiliation{Van der Waals-Zeeman Institute, University of Amsterdam, 1018XE Amsterdam, The Netherlands}

\author{D.~van der Marel}
\affiliation{D\'{e}partement de Physique de la Mati\`{e}re Condens\'{e}e, Universit\'{e} de Gen\`{e}ve, CH-1211 Gen\`{e}ve 4, Switzerland}

\date{\today}


\begin{abstract}

We performed far-infrared optical spectroscopy measurements on the heavy fermion compound URu$_2$Si$_2$ as a function of temperature. The light's electric-field was applied along the $\mathbf{a}$-axis or the $\mathbf{c}$-axis of  the tetragonal structure. We show that in addition to a pronounced anisotropy, the optical conductivity exhibits for both axis a partial suppression of spectral weight around 12~meV and below 30~K. We attribute these observations to a change in the bandstructure below 30~K. However, since these changes have no noticeable impact on the entropy nor on the DC transport properties, we suggest that this is a crossover phenomenon rather than a thermodynamic phase transition.

\end{abstract}

 \maketitle



\section{INTRODUCTION}

For more than two decades the heavy fermion compound URu$_2$Si$_2$ has attracted special interest due to a very rich phase diagram as function of temperature, magnetic field and pressure~\cite{Kim03,Villaume08}. In particular, as a function of temperature, two clear jumps were observed by specific heat~\cite{Maple86}. One occurs at $\sim$~1.5~K and is associated to a superconducting transition, the other takes place at T$_{HO}$~=~17.5~K and is attributed to a hidden order (HO) transition whose order parameter is still unidentified despite numerous theoretical and experimental efforts.  For many years, the order parameter has been associated with a small antiferromagnetic moment (SMAF) of $\sim$~0.03~$\mu_B$ per U atom observed below T$_{HO}$ \cite{Broholm87}, but recent NMR~\cite{Matsuda01} and $\mu$-SR~\cite{Amitsuka02} measurements revealed that this magnetic moment is strongly sample dependent and disappears in stress-free samples~\cite{Amitsuka07,Yokoyama05}.

It is believed that the HO transition is concomitant with a reconstruction of the Fermi surface (FS). This is supported for instance by specific heat~\cite{Maple86}, resistivity~\cite{Maple86,Palstra86} and inelastic neutron scattering (INS)~\cite{Wiebe07,Janik09,Bourdarot10}. The exponential behavior of the two first quantities below T$_{HO}$ clearly indicates the opening of a partial gap in the density of states whose amplitude is of about 10~meV. Finally, at low temperature, URu$_2$Si$_2$ is a compensated semimetal with a low carrier density ($\sim6\times10^{20}$~cm$^{-3}$) but with a high mobility~\cite{Levallois09,Behnia05,Bel04}.

The physics of URu$_2$Si$_2$ is shaped by two main ingredients: itinerant $s$-$d$ electrons and more localized magnetic 5$f$ electrons. Below the coherence temperature T$_{coh}\simeq$70~K: the conduction electrons hybridize with the more localized magnetic $f$-electrons and experience an important mass enhancement. Here the magnetic moments of the $f$ atoms tend to form a coherent state and the renormalized conduction electrons behave as a Fermi liquid. For the HO phase, recent theories point to a dual itinerant and localized description and propose a wide variety of order parameters such as spin or charge density waves~\cite{Ikeda98,Mineev05,Maple86}, orbital antiferromagnetism~\cite{Chandra02} or multipolar orders~\cite{Santini94,Kiss05,Ohkawa99,Haule09,Harima10}. Recent fully itinerant band structure calculations propose that symmetry breaking induced by a dynamic dynamic mode of AF moment excitations cause the HO gap~\cite{Elgazzar09}. Furthermore, spin nematic~\cite{Pepin2011} and hybridization wave~\cite{Dubi2011} order parameters have also been proposed. In this context, far-infrared optical spectroscopy  is a powerful tool to explore how the itinerant $s$-$d$ electrons and the localized magnetic $f$-electrons interact in URu$_2$Si$_2$, as it probes the spectral distribution of the charge carriers.

Here we present far-infrared reflectivity measurements with the electric component of the electromagnetic field polarized along the $\mathbf{a}$-axis and the $\mathbf{c}$-axis of URu$_2$Si$_2$'s tetragonal structure. In addition to a pronounced anisotropy we observe a partial suppression of the optical conductivity for both axis at around 12~meV and below $\sim$~30~K. This behavior appears to be a precursor of the HO phase transition at 17.5 K.


\section{EXPERIMENTAL}

The single crystal was grown with the traveling floating-zone technique at the Amsterdam/Leiden Center~\cite{Niklowitz09}. The DC resistivity of the sample clearly exhibits the HO transition at 17.5~K. The crystal was oriented by X-ray Laue diffraction, cut with a diamond saw along the $\mathbf{ac}$-plane and polished. The reflectivity was measured for light propagating perpendicular to this plane. The direction of the light's electric field $\mathbf{E}$ was selected by a grid-wire gold polarizer with an accuracy better than 2$^o$ along the $\mathbf{a}$- or $\mathbf{c}$-axis. Measurements were performed combining infrared reflectivity (30 - 7000~cm$^{-1}$) and ellipsometry (6000 - 45000~cm$^{-1}$)  (1~eV = 8065.8~cm$^{-1}$). The sample was mounted in a helium flow high vacuum cryostat. The cryostat design is optimized to obtain very high position stability of the sample and permits measurement of details of the temperature dependence of the reflectivity such as shown in Fig.~\ref{RT}. The price one pays is that the lowest temperature reached is $\sim$~10~K. In situ gold evaporation on the sample allowed us to extract the absolute value of the reflectivity by computing the ratio of the signal measured on the sample by the signal measured on the gold layer. Two different ways of scanning temperature were used: (i) various stabilized temperatures, where reflectivity spectra were measured at 11 temperatures between 15~K and 280~K stabilized within $\pm$~0.5 K and (ii) continuous sweeping down and up of the temperature, where the constant rate was such that one measurement corresponds to a temperature interval of 1 to 2~K. The reflectivity obtained with the two types of acquisition match exactly for the corresponding temperatures. The thickness of the sample ($\sim$~4~mm) caused a difference in temperature of $\sim$~5~K, at the lowest temperature reached, between the surface of the sample and the cold finger. This difference was calibrated by comparing the DC resistivity with the zero frequency limit of 1/$\sigma_1(\omega)$.

\section{RESULTS}


\begin{figure}[t]
\centering
\includegraphics[width=9cm]{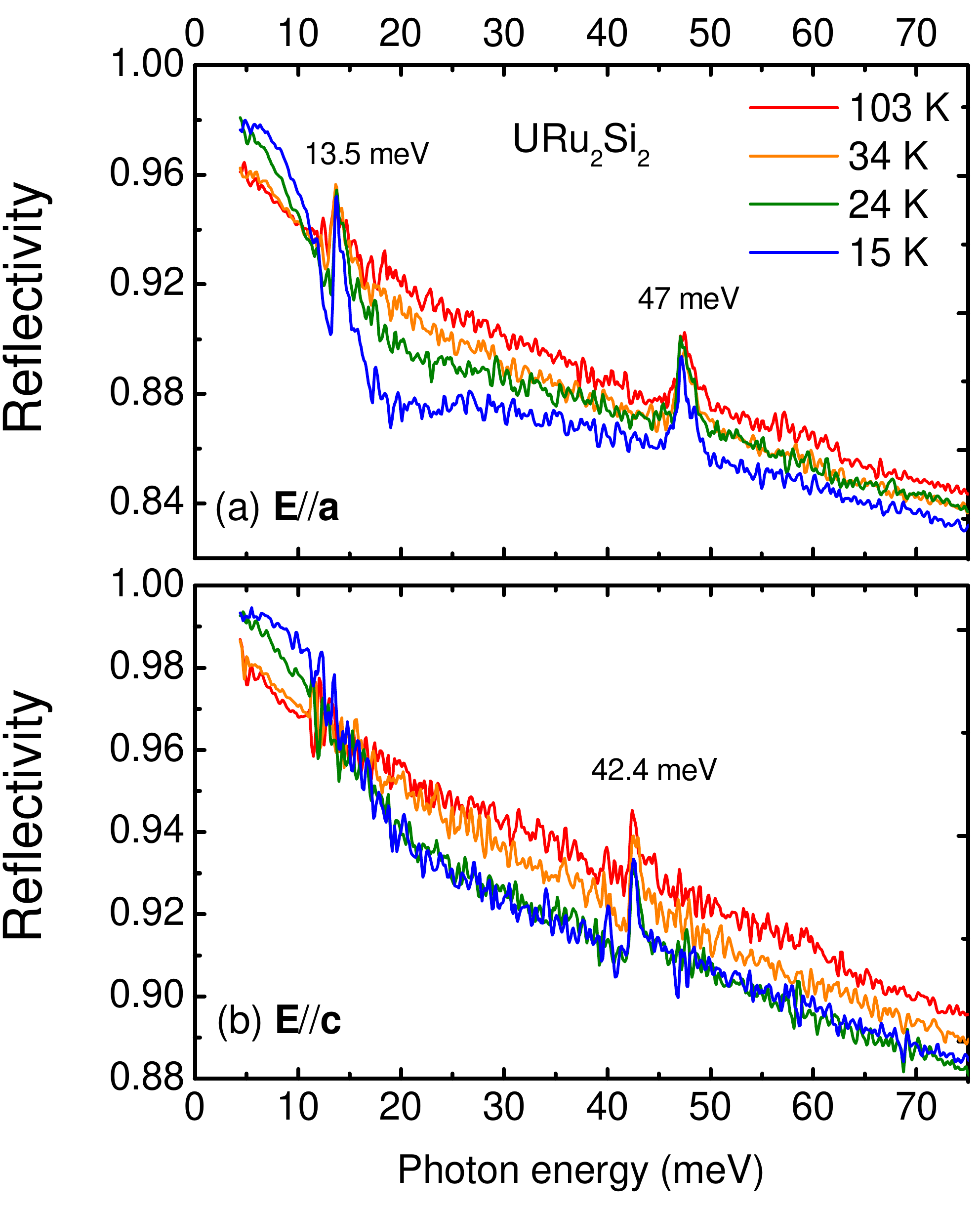}\\
\caption{(Color online) \textbf{Optical far-infrared reflectivity along the $\mathbf{a}$- and the $\mathbf{c}$-axis} as a function of frequency for different temperatures. The reflectivity is anisotropic. Two phonons are observed along the $\mathbf{a}$-axis at 13.5~meV (109~cm$^{-1}$) and 47~meV (380~cm$^{-1}$), one is observed along the $\mathbf{c}$-axis at 42.4~meV (342~cm$^{-1}$).}
\label{reflectivity}
\end{figure}

The far-infrared optical reflectivity of URu$_2$Si$_2$ for electric field applied along the $\mathbf{a}$- or the $\mathbf{c}$-axis is displayed in panel (a) and (b) of Fig.~\ref{reflectivity} for four different temperatures. The reflectivity measured along the $\mathbf{a}$-axis is consistent with data at the same temperatures reported by Bonn~\textit{et al.}~\cite{Bonn88}. There is a pronounced anisotropy between the $\mathbf{a}$- and $\mathbf{c}$-axis. First, as anticipated from resistivity measurements, the $\mathbf{c}$-axis is the most metallic one with a higher value of reflectivity. Second, two phonons are observed along the $\mathbf{a}$-axis at 13.5~meV (109~cm$^{-1}$) and 47~meV (380~cm$^{-1}$), while a third different one is observed along the $\mathbf{c}$-axis at 42.4~meV (342~cm$^{-1}$). The positions of the phonons do not show any sizeable temperature dependence. In contrast, a minimum develops near 20~meV as the temperature decreases. This minimum is more pronounced along the $\mathbf{a}$-axis and is  already clearly distinguishable in the 24~K spectrum, which is well \textit{above} the HO temperature T$_{HO}$~=~17.5~K. As the minimum develops, the reflectivity strongly increases at low frequency. One would expect the reflectivity at low frequency to decrease if the minimum were caused by a spin- or charge density wave~\cite{Dressel}. The increase of reflectivity signals that part of the FS remains ungapped, and that scattering is suppressed at the same time that the minimum develops.
A dip in the reflectivity at 5~meV (42~cm$^{-1}$) develops below below 17.5~K in the $\mathbf{a}$-axis reflectivity~\cite{Bonn88,Degiorgi96,Thieme96,Lobo10}. We have also observed this feature, which is just below the temperature and frequency ranges displayed in Fig.~\ref{reflectivity}, on the same crystal and the same geometry of electromagnetic field and sample surface in a separate setup where the sample was immersed in He contact gas for temperatures down to 5~K.


\begin{figure}[t]
\centering
\includegraphics[width=8cm]{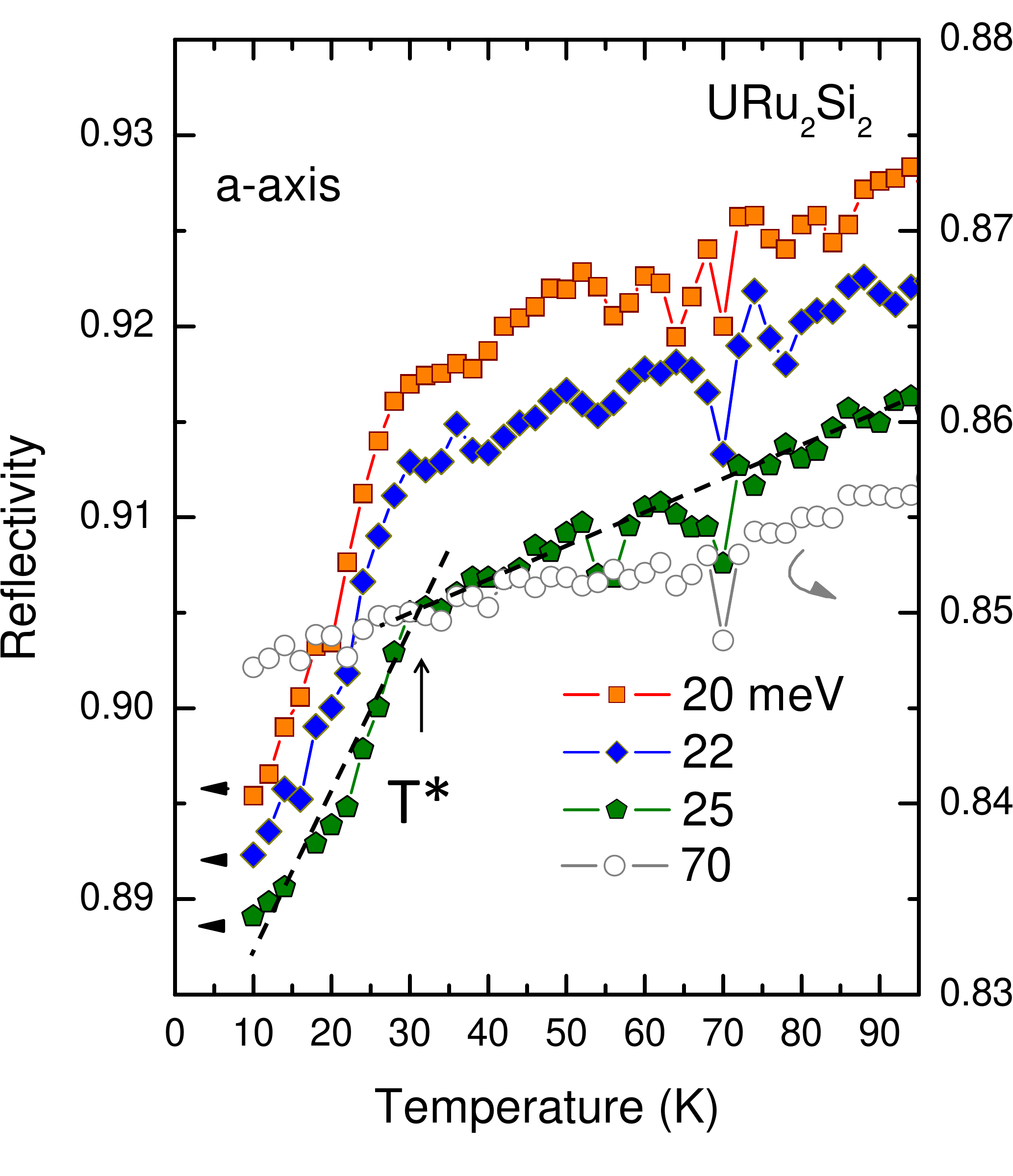}
\caption{(Color online) \textbf{Temperature dependence of the $\mathbf{a}$-axis reflectivity} for selected energies. The temperature T$^{\star}$ of the crossover behavior is determined by the crossing of the two straight dotted lines.}
\label{RT}
\end{figure}


The temperature dependence of the $\mathbf{a}$-axis reflectivity for selected energies is presented in Fig.~\ref{RT}. Cooling down to 30~K the reflectivity presents a slow and continuous decrease with temperature in the energy range 20-30~meV. However, for frequencies between $\sim 10$ and $\sim 30$ meV the reflectivity exhibits a noticeable drop of almost 2$\%$ when the temperature is lowered below T$^{\star}\approx 30$ K, indicating a significant change of the electronic properties of URu$_2$Si$_2$ around this temperature. The abrupt change of temperature dependence at 30~K was confirmed by repeating the same measurement several times. We obtained the same temperature dependence for measurements sweeping the temperature up and sweeping it down, ruling out thermalization issues. The reflectivity along the $\mathbf{c}$-axis (not shown) presents the same behavior but with a smaller amplitude of the change below T$^{\star}\simeq$30~K.


\begin{figure}[t]
\centering
\includegraphics[width=9cm]{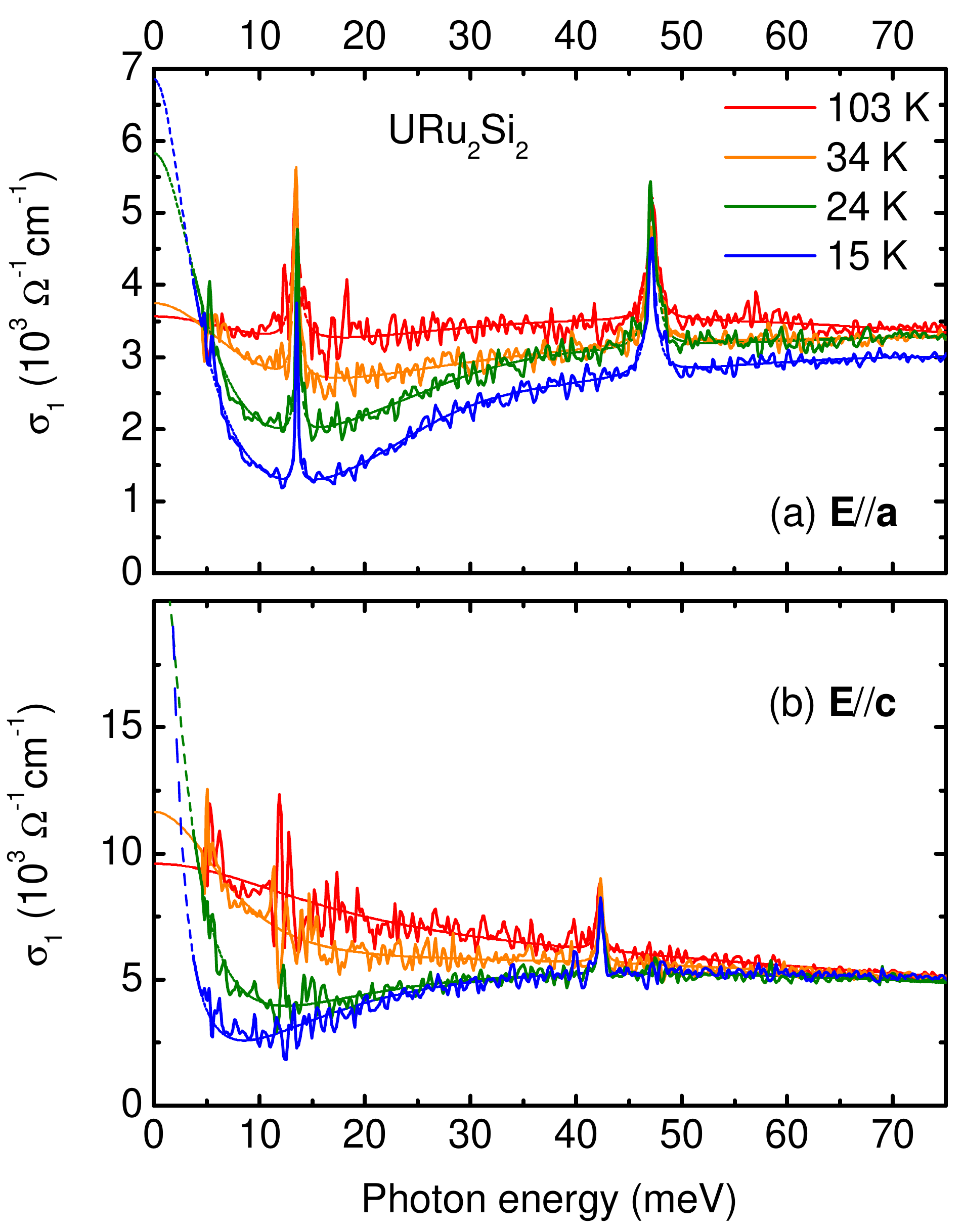}
\caption{(Color online) \textbf{Real part of the optical conductivity along the $\mathbf{a}$- and the $\mathbf{c}$-axis as a function of frequency} at different temperatures. The thin dashed lines which extend to zero frequency are Drude-Lorentz fits. The measurements were performed down to ~4~meV.}
\label{conductivity}
\end{figure}

The optical conductivity shown in Fig.~\ref{conductivity} for electric field along the $\mathbf{a}$- and the $\mathbf{c}$-axis was determined using the Kramers-Kronig relations. The thin dashed lines which extend to zero frequency are Drude-Lorentz fits
\begin{equation}
\sigma(\omega) = \frac{\omega}{4\pi i} \sum_j\frac{\omega_{p,j}^2}{\omega_{0,j}^2-\omega^2-i\gamma_j\omega}
\label{Drude_Lorentz}
\end{equation}
where $\omega_{0,j}$ the transverse frequency, $\gamma_j=1/\tau_j$ is the inverse lifetime and $\omega_{p,j}^2$ the spectral weight of the $j$-th Lorentz oscillator.
As the temperature is decreased, a clear minimum in $\sigma_1(\omega)$ develops around 15~meV along the  $\mathbf{a}$-axis. A similar, but less marked minimum occurs along the $\mathbf{c}$-axis near 11~meV. Accompanying this loss of spectral weight (SW), there is a development and a narrowing of the Drude peak, {\em i.e.} the peak centered at zero frequency which corresponds to intra-band processes at the Fermi level ($\varepsilon_F$) and thus is the free carrier signature. The width of this peak is proportional to the scattering rate $1/\tau$ of the free carriers. Consequently, given that the effective mass is smaller in the HO phase than in the "normal" state as deduced from specific heat measurement~\cite{Maple86}, the fact that the peak narrows indicates that the carriers become more mobile.

\begin{figure}[t]
\centering
\includegraphics[width=9cm]{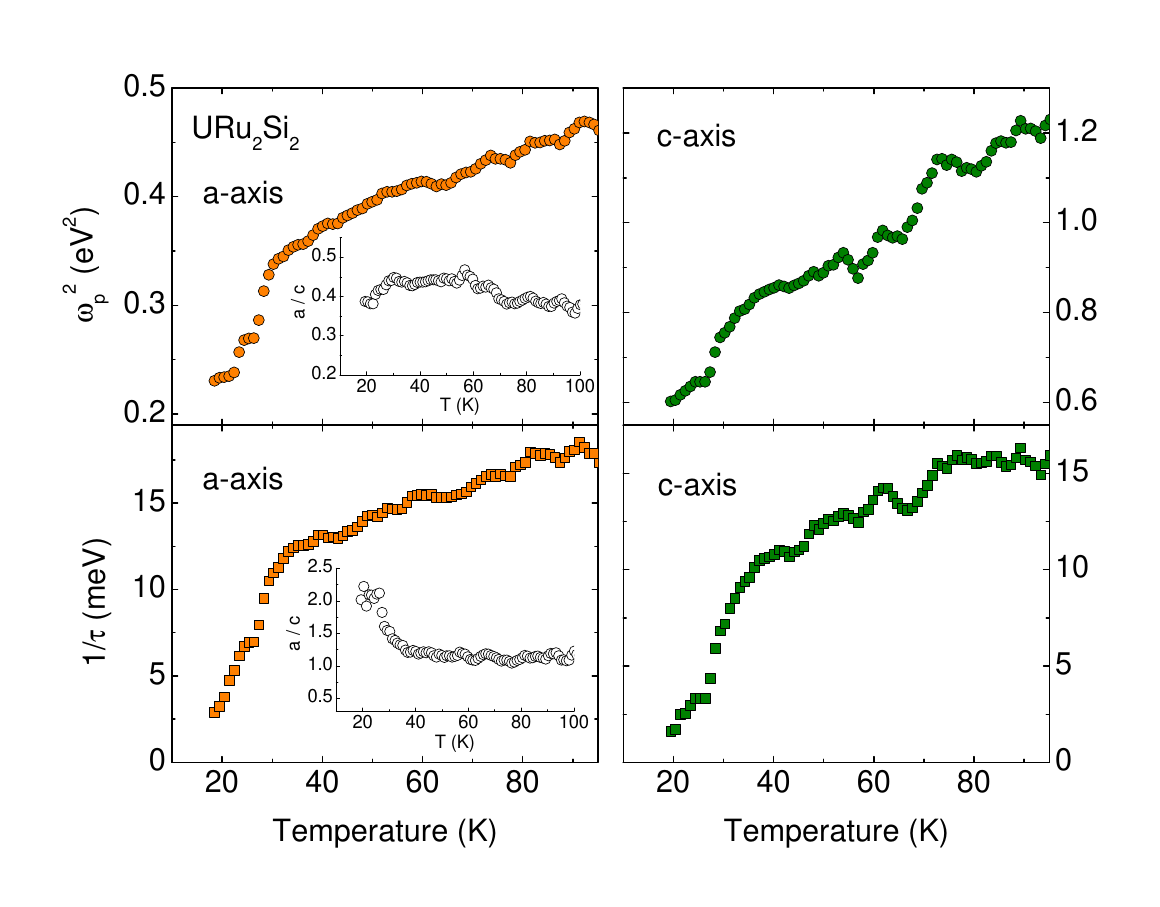}
\caption{(Color online) \textbf{Free carrier spectral weights (up panels) and relaxation rates (bottom panels) along the $\mathbf{a}$- and the $\mathbf{c}$-axis as a function of temperature} extracting from the Drude-Lorentz fits. The insets present the ratio of each quantity between $\mathbf{a}$- and $\mathbf{c}$-axis.}
\label{aniso}
\end{figure}

The coherent part of the free carrier optical conductivity is given by the $j=0$ component in Eq.~\ref{Drude_Lorentz} with $\omega_{0}=0$, the corresponding spectral weight of which is given by the fitted $\omega_p^2$. This procedure separates the coherent part of the free carrier response from finite frequency modes partly overlapping with the zero-frequency mode. The fitted $\omega_p^2$ are displayed in Fig.~\ref{aniso} for both $\mathbf{a}$- and $\mathbf{c}$-axis together with the corresponding relaxation rates 1/$\tau$. Along both axes a sharp drop of $\omega_p^2$ and 1/$\tau$ is observed at a temperature T$^{\star}\sim~$30~K, almost two times T$_{HO}$, indicating that the electronic structure begins to change. Accordingly, our analysis indicates that both the width of the Drude peak and its area reduce below T$^{\star}$ along both axes, which strongly points to a large decrease of the carrier density. Hence, surprisingly, at a temperature almost two times T$_{HO}$ the electronic structure begins to change and a partial gap develops around 10~meV. The anisotropy of the spectral weights shown in the inset of left upper panel in Fig.~\ref{aniso} amounts to a factor of $\sim$~0.4 for T~$>$~90~K, and does not change significantly as the temperature is swept though 30~K. Turning now to the relaxation rates, we see from the inset of the left lower panel in Fig.~\ref{aniso} that the relaxation rate is isotropic above 30~K, but becomes progressively more anisotropic ($\tau^{-1}_a$ / $\tau^{-1}_c \sim$ 2) below T$^{\star}$.



\begin{figure}[t]
\centering
\includegraphics[width=7cm]{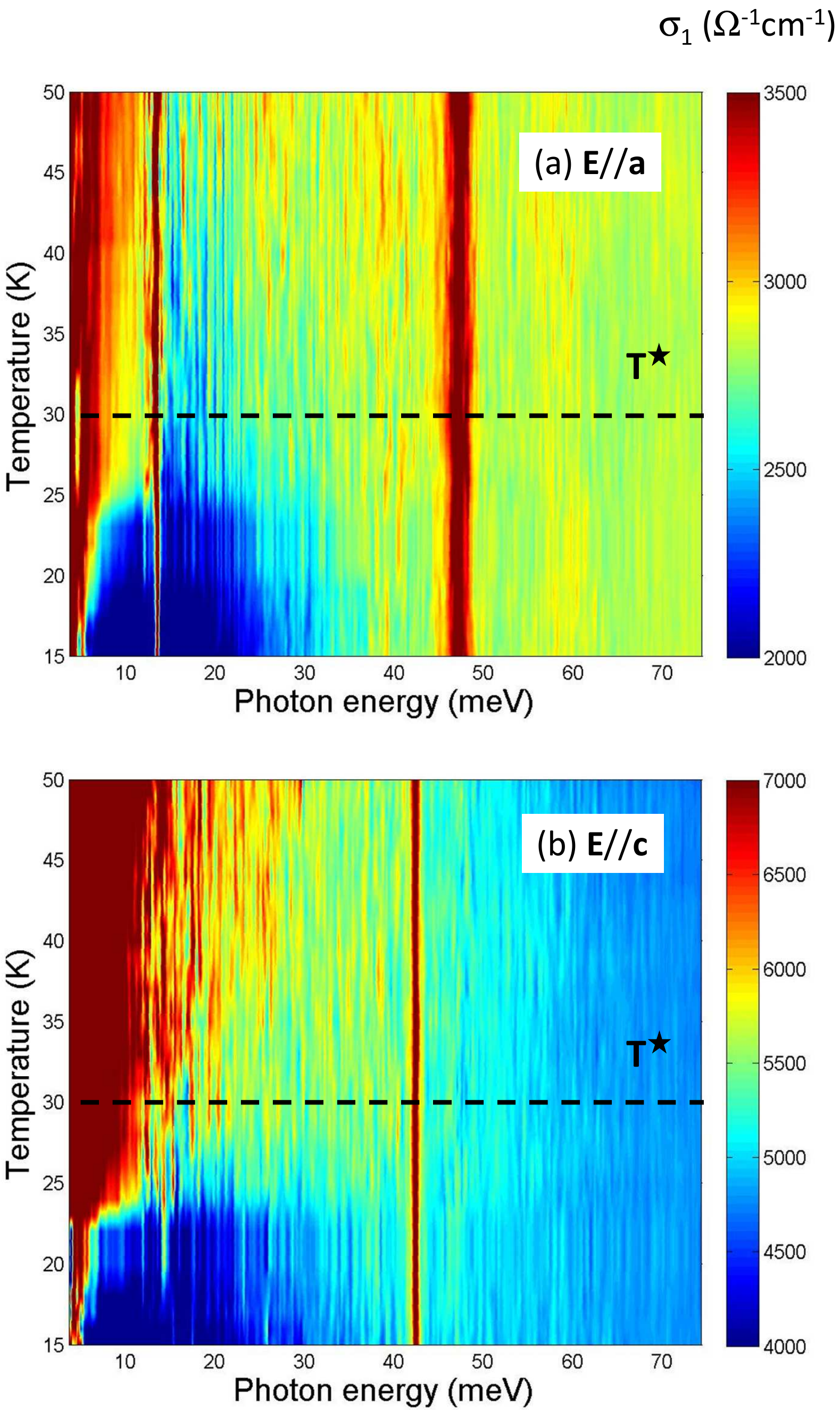}
\caption{(Color online) \textbf{Temperature-frequency colormap of the conductivity along the $\mathbf{a}$- and the $\mathbf{c}$-axis}. The blue areas correspond to the loss of spectral weight as the temperature decreases. The dashed lines indicate the pseudogap crossover temperature T$^{\star}$.}
\label{colorplot}
\end{figure}

In Fig.~\ref{colorplot}, the temperature-frequency colormap of the conductivity $\sigma_1(\omega)$ above T$_{HO}$ along $\mathbf{a}$- and $\mathbf{c}$-axis highlights in blue the loss of SW that develops as the temperature decreases. The significant loss of SW starts at around T$^{\star}\simeq$30~K, {\em i.e.} far above the HO temperature T$_{HO}$~=~17.5~K. The loss of SW is maximum at a frequency which is slightly different along the two axis and equals $\simeq$~15~meV (120~cm$^{-1}$) along the $\mathbf{a}$-axis and $\simeq$~11~meV (90~cm$^{-1}$) along the $\mathbf{c}$-axis, which corresponds to the 10~meV gap estimated by specific heat at the HO transition. The sudden suppression of optical conductivity takes place far below T$_{coh}\simeq$70~K, and the question whether or not there is a relation to the hybridization gap opening at T$_{coh}$, needs to be examined carefully. For the purpose of this discussion we reserve the label "pseudogap" for the suppression of optical conductivity, {\em i.e.} an incomplete suppression of density of states not necessarily due to a spontaneous symmetry breaking and/or a phase transition~\cite{Mott68}.


\begin{figure}[t]
\centering
\includegraphics[width=9cm]{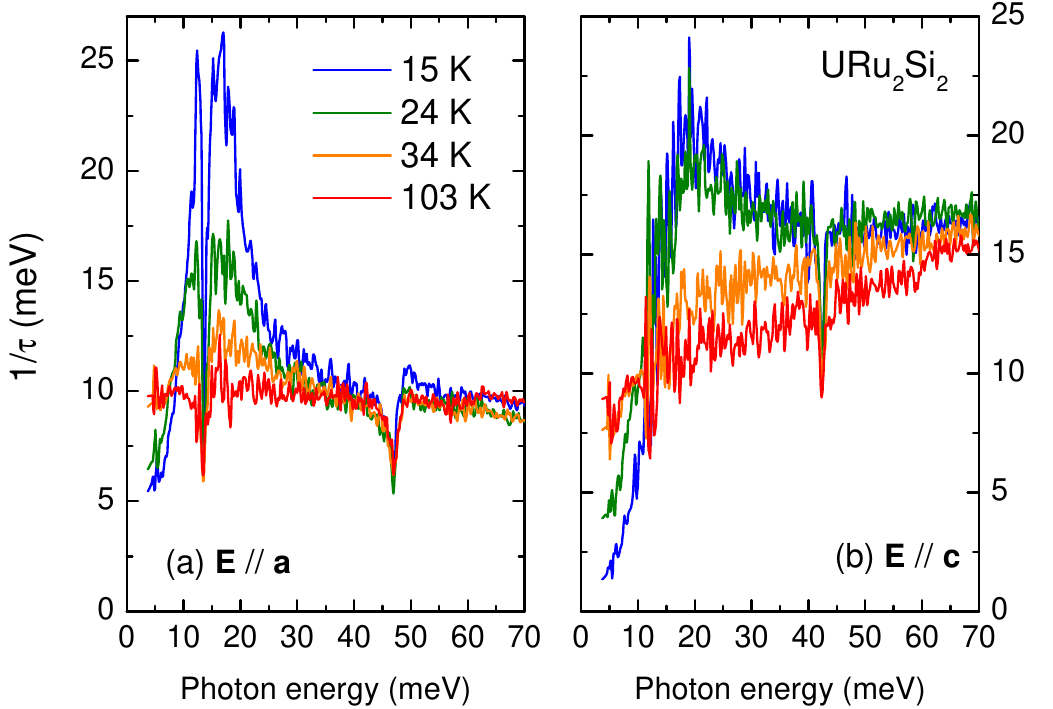}
\caption{(Color online) \textbf{Frequency dependent scattering rate along the $\mathbf{a}$- and the $\mathbf{c}$-axis} at different temperatures, calculated according to the extended Drude formalism. A maximum in 1/$\tau(\omega)$ along both axis below 30~K is observed, which becomes increasingly pronounced at lower temperatures.}
\label{invtau}
\end{figure}

The frequency dependence of the scattering rate defined as $1/\tau(\omega)=\frac{\omega_p^2}{4\pi}Re(\frac{1}{\sigma(\omega)}$) for both axes, obtained according to the extended Drude formalism~\cite{Allen77}, is shown in Fig.~\ref{invtau}. We used the plasma frequencies $\omega_p$ = 496~meV and $\omega_p$ = 793~meV for $\mathbf{a}$- and $\mathbf{c}$-axis respectively, according to the Drude weight measured at 20~K (see Fig.~\ref{aniso}). It is important to note that, while the absolute value of $1/\tau(\omega)$ depends on the choice of $\omega_p$, the shape of the spectra is robust. One can see that at 100~K, the scattering rate is almost constant for the $\mathbf{a}$-axis whereas it increases monotonically with frequency for the $\mathbf{c}$-axis. Nevertheless, little occurs as a function of decreasing temperature until T$^{\star}\simeq$~30~K is reached. Contrary to what is normally expected, one observes that below $\sim$30~K, there is a maximum in $1/\tau(\omega)$ along both axis, which becomes increasingly pronounced at lower temperatures. This maximum of $1/\tau(\omega)$ has already been observed by Bonn {\em et al.}\cite{Bonn88} and was confirmed for 20, 30 and 50~K in a recent preprint by Nagel {\em et al.}\cite{Nagel11}. If $1/\tau(\omega)$ is generated by inelastic processes, it should be a monotonically increasing as a function of $\omega$. The maximum therefore reveals the presence of interband transitions at $\sim$15~meV. The fact that this intensity develops as temperature is lowered below 100~K, suggests that this are transitions across the hybridization gap which opens at T$_{coh}$. Moreover, we observe a drop of $1/\tau(\omega)$ when the temperature is lowered below T$^{\star}=30$~K for frequencies in the limit $\omega\rightarrow$~0, which could be attributed to an additional frequency dependent scattering at very low frequencies below 4~meV. This is not unexpected when the Landau quasi-particles are coupled to a reservoir of low energy collective excitations.
%


\section{DISCUSSION}

The main result of this work concerns the decrease of the SW below T$^{\star}\simeq$30~K,
approximately twice T$_{HO}$ and half T$_{coh}$ ($\simeq$70~K).
%
%
This is not the first observation of a noticeable phenomenon well above T$_{HO}$: (i) The infrared reflectivity at 20~K of Bonn~\textit{et al.}~ have in fact a weak minimum at $\sim$~10~meV (Fig.1 of Ref.~\onlinecite{Bonn88}). (ii) A $\sim$~10~meV gap was observed at $\simeq$~22~K by point-contact spectroscopy~\cite{Hasselbach92,Rodrigo97}. Using the same technique Park {\em et al.}\cite{Park11} observed the opening of a 12~meV gap at T$\sim$30~K, in close agreement with our optical data. (iii) Recent inelastic neutron scattering data reveal a quasi-elastic magnetic continuum due to spin fluctuations which persists up to $\sim$~30~K~\cite{Bourdarot10}.
Whereas the DC resistivity and the specific heat reveal a clean second order phase transition at T$_{HO}$, these probes show a smooth temperature dependence at T$^{\star}$, unlike our infrared data.

Since the opening of an optical pseudogap is not accompanied by an abrupt change in the DC resistivity, we  conclude that either states at the Fermi level are not affected at all, or that the affected states contribute little to the DC resistivity. The latter scenario would imply that parts of the Fermi surface which become gapped have a very low mobility. The published specific heat~\cite{vanDijk97} data confirm this: if anything at all happens near 30~K, it would have to be a suppression of C$_V$/T. This excludes scenarios whereby the effective mass increases below 30~K, and instead points toward a possible weak suppression of density of states at the Fermi level. The latter is also supported by by point-contact spectroscopy data, indicating a partial reduction of density of states at $\varepsilon_F$ below $\sim$30~K~\cite{Park11}.

Above 100~K the magnetic susceptibility can be explained by a model of 5$f$ moments. Due to the intersite coupling, these moments are expected to be transformed into a partially filled narrow band at a sufficiently low temperature. An important question is, to what extent the 5$f$ electrons are localized or itinerant for T$\sim$T$_{HO}$. The model of Haule and Kotliar~\cite{Haule09} based on crystal-field split localized $5f$-multiplets coupled to a bath of conduction electrons was supported by low temperature scanning tunneling microscopy data\cite{schmidt2010,aynajian2010} at and below T$_{HO}$. However, recent observations~\cite{Janik09} of highly dispersive paramagnons indicate an itinerant form of magnetism for temperatures above T$_{HO}$. This supports the notion that the transformation of 5$f$ moments to narrow bands takes place at the crossover temperature T$_{coh}\sim$ 70~K, opening a gap at $\varepsilon_F$ due to hybridization with a partially filled wide band. This is believed to give rise to a hybridization gap, and is strongly supported by the fact that $1/\tau(\omega)$ exhibits a maximum near 15~meV. The latter is compatible with a doubling of the unit cell along the $\mathbf{c}$-direction, as observed under pressure in the antiferromagnetic phase~\cite{Villaume08}. Such doubling in the HO phase has been predicted~\cite{Harima10,Elgazzar09} based on temperature dependence of a narrow band close to the Fermi energy seen in angle resolved photoemission  experiments~\cite{Santander09,Yoshida2010}, and on reentrant behavior of the HO in a magnetic field~\cite{Aoki09}.

The question remains: what happens at -and below- T$^{\star}$ = 30~K? The hybridization gap has already been formed at the much higher T$_{coh}$. It appears that part of the Fermi surface becomes depleted in a rather gradual way due to a fluctuation gap, such as for example has been proposed in Ref.~\onlinecite{Haralsden11}. Since the specific heat exhibits a clean second order drop at T$_{HO}$ without a fluctuation tail extending to high temperature, the amplitude fluctuations of the hidden order gap are an unlikely candidate for the suppression of the optical conductivity.

\section{CONCLUSION}

In conclusion, we have shown that in addition to a pronounced anisotropy, the optical conductivity of URu$_2$Si$_2$ exhibits a clear suppression of spectral weight around 15~meV for the $\mathbf{a}$-axis and 11~meV for the $\mathbf{c}$-axis. Most remarkably, we found that a partial gap emerges at 30~K, well above the hidden order temperature T$_{HO}$~=~17.5~K and whose amplitude seems to be T-independent and not affected by the hidden order transition. As this suppression coincides with the development, at the same energy, of a maximum in the scattering rate, indicating optical interband transitions, we propose that this partial gap is a hybridization gap. We suggest that this electronic reconfiguration below 30~K is a precursor of the hidden order state.

\begin{acknowledgements}
The authors thank A. Balatsky, L. Greene, A. Kuzmenko and R. Lobo for helpful discussions. This work has been supported by the SNSF through Grant No. 200020-135085 and the National Center of Competence in Research (NCCR) "Materials with Novel Electronic Properties-MaNEP".
\end{acknowledgements}



\end{document}